\documentclass[sigconf]{acmart}

\AtBeginDocument{%
  \providecommand\BibTeX{{%
    \normalfont B\kern-0.5em{\scshape i\kern-0.25em b}\kern-0.8em\TeX}}}

\copyrightyear{2023}
\acmYear{2023}
\setcopyright{acmlicensed}
\acmConference[L@S '23] {Proceedings of the Tenth ACM Conference on Learning @ Scale}{July 20--22, 2023}{Copenhagen, Denmark.}
\acmBooktitle{Proceedings of the Tenth ACM Conference on Learning @ Scale (L@S '23), July 20--22, 2023, Copenhagen, Denmark}
\acmPrice{15.00}
\acmISBN{979-8-4007-0025-5/23/07}
\acmDOI{10.1145/3573051.3593382}

\usepackage{subfig}
\begin{document}

\title{Evaluating a Learned Admission-Prediction Model as a Replacement for Standardized Tests in College Admissions}


\author{Hansol Lee}
\affiliation{%
  \institution{Stanford University}
  \city{Stanford}
  \state{CA}
  \country{USA}}
\email{hansol@stanford.edu}

\author{Ren\'e F. Kizilcec}
\affiliation{%
  \institution{Cornell University}
  \city{Ithaca, NY}
  \country{USA}}
\email{kizilcec@cornell.edu}

\author{Thorsten Joachims}
\affiliation{%
  \institution{Cornell University}
  \city{Ithaca, NY}
  \country{USA}}
\email{tj@cs.cornell.edu}

\begin{abstract}
A growing number of college applications has presented an annual challenge for college admissions in the United States. Admission offices have historically relied on standardized test scores to organize large applicant pools into viable subsets for review. However, this approach may be subject to bias in test scores and selection bias in test-taking with recent trends toward test-optional admission. We explore a machine learning-based approach to replace the role of standardized tests in subset generation while taking into account a wide range of factors extracted from student applications to support a more holistic review. We evaluate the approach on data from an undergraduate admission office at a selective US institution (13,248 applications). We find that a prediction model trained on past admission data outperforms an SAT-based heuristic and matches the demographic composition of the last admitted class. We discuss the risks and opportunities for how such a learned model could be leveraged to support human decision-making in college admissions.
\end{abstract}





\maketitle
\section{Introduction}
Colleges and universities across the United States receive an increasing number of student applications for admission to their incoming class each year~\cite{nietzel_2021}. The \textit{Common Application} is the primary tool that applicants use to apply to colleges in the United States.\footnote{https://www.commonapp.org/} It received over 6.6 million first-year applications during the 2021-22 admission cycle, which constitutes a 9.1\% increase over the previous year, and a 21.3\% increase over the 2019-20 cycle~\cite{commonapp}. The growing number of applications received by selective colleges and universities has presented an annual challenge for college admission, especially for institutions that follow a holistic review process. Holistic review aims to assess each applicant as a whole by considering a wide range of factors presented in a student’s application, which makes it a thorough but time-consuming process~\cite{Espinosa2015RaceCA, whatarewetalking, Stevens+2020}. 


The large volume of applications received by colleges challenges human reviewers to perform a thoughtful and equitable review of individual applications given the limited admission timeline, which typically spans only a few months. 
In order to manage the reviewing load, admission officers lean on quantitative measures to prioritize the order of human review and to most effectively allocate the limited reviewing resources. This has led many selective colleges in the United States to use heuristics that are based on standardized test scores, such as the ACT or SAT, to ``triage'' their large applicant pool---that is, to organize their large applicant pool to better allocate the limited resources available for application review~\cite{princeton}. However, there are several limitations to using standardized test scores to triage the applicant pool, a practice we will refer to as the ``SAT-based'' approach. First, there are many unresolved concerns about gender, racial, and socioeconomic biases in standardized test scores which may undermine the fairness of the SAT-based approach for organizing the applicant pool \cite{freedle, doi:10.1177/003172170208400411, doi:10.1177/016146811311500406, rosser1989sat, gender_sat, doi:10.3102/0013189X18762105}. Second, this SAT-based approach is dependent on requiring all applicants to submit their test scores, which may impose a significant financial burden on many applicants. Moreover, many institutions began test-optional admission in response to testing site closures during the coronavirus disease (COVID-19) pandemic, which makes the traditional SAT-based heuristics incomplete, impractical, and potentially subject to selection biases~\cite{bennett2022untested}.

Admission offices may consider alternatives to the SAT-based approach in order to overcome the issue of bias and costliness of standardized tests, their increasing unavailability in test-optional admission, and their non-holistic nature as a basis for organizing the admission process. To this end, we explore a machine learning-based approach that aims to mimic holistic review by taking into account a wide range of factors extracted from student applications---not just standardized test scores---to predict whether or not an applicant will be admitted. In particular, we focus on answering the following research question: How well can an admission prediction model trained on past admission data replace and improve on the traditional SAT-based heuristic to organize the applicant pool for review? We examine this question in the context of first-year undergraduate admissions at a selective US institution and find that the prediction model is better aligned with existing admission practices at the case institution compared with an SAT-based heuristic.

\section{Related Work}
Our work builds on a framework of human-machine collaboration, which advocates for the design and use of machine learning systems with the intention of augmenting, not replacing, human contributions~\cite{miller, wilson2018collaborative}. Autor and colleagues have argued that machines may replace humans in performing routine tasks while complementing humans in performing non-routine cognitive tasks~\cite{10.2307/25053940}. Jarrahi suggested that machines may extend human cognition by equipping human decision-makers with comprehensive data analytics, whereas humans may offer a more holistic and intuitive approach to decision-making \cite{JARRAHI2018577}. In our work, we aim to leverage the complementary strengths of machines and humans in the admission process; a machine-learned admission prediction model can be used to organize a large applicant pool in a way consistent with past institutional decision-making, allowing admission officers to use the freed-up resources to engage in the process of holistic review in a more meaningful way.

Approaches based on machine learning are increasingly studied to support various aspects of college admissions. For example, Basu and colleagues used machine learning algorithms to predict whether a student who is offered admission would accept the offer~\cite{data4020065}. This is helpful for institutions that need an accurate estimate of the size of their entering class. From the applicants’ perspective, machine learning techniques can be used to evaluate applicants' chances of admission to help students make informed decisions about where to apply to college~\cite{7836726, kiaghadi}. In the context of college essays, Alvero and colleagues explored the use of computational text analysis to assist human readers in their evaluation of college application essays~\cite{alvero2020ai, Alvero2022-im}. 

Many studies have examined the use of machine learning to predict admission outcomes to support the holistic review process of admission officers. In the 1990s, Bruggink et al.\ and Moore et al.\ utilized domain knowledge to build statistical models to predict undergraduate admission decisions~\cite{10.2307/40196173, MOORE1998659}. More recently, Lux and colleagues used multi-layer perceptron and support vector machine algorithms to predict admission decisions at a small private liberal arts college~\cite{lux}. Rees and Ryder evaluated the usefulness of the random forest algorithm in assisting in the process of an internal medicine residency program in northern New England~\cite{rees}. Neda and Gago-Masague compared classification performances of various machine learning algorithms trained on applications submitted to the Computer Science department at the University of California, Irvine~\cite{neda}. 

However, most prior work evaluated the learned admission-prediction model by reporting the overall model performance without explicit discussion of the implications of their use in admission practice. For instance, one recent study examined the feasibility of machine learning support for undergraduate admissions by reporting an overall accuracy of 0.783 and an AUROC of 0.871~\cite{neda}. However, such numbers alone may not be sufficient to inform how the learned prediction model should be used in practice. They often lack a meaningful baseline model performance to compare with. Additionally, they may not provide enough insights about the limitations of the prediction model. Our work utilizes a similar modeling approach to estimate admission probability for each applicant but provides additional analyses of the model predictions with direct comparison to the SAT-based approach as a baseline. We also provide specific recommendations for using the predicted admission probabilities in practice in order to avoid potential misuse of the prediction model in a superficial nature~\cite{kizilcec2020algorithmic}.

Among the prior work that involved the use of machine learning for predicting admission decisions, Waters and Miikkulainen examined a contentious use of an admission prediction model: to save time in the admission process~\cite{grade, grade_critique}. In their 2014 study, Waters and Miikkulainen used logistic regression to predict graduate admissions in the computer science department at the University of Texas at Austin. Their work focused on improving the efficiency of review by cutting the time spent on application reading and the number of applications to review. We emphasize that our work focuses on improving the process of holistic review by providing a tool that better supports the organization of the applicant pool than the SAT-based method previously used in admission, not by providing a shortlisting tool to make the admission process cost-effective.
\section{Methods}
In this section, we describe the first-year undergraduate admission process at the case university and explain how we develop and evaluate the admission prediction model using past admission data at the case university. We then explore how to construct new applicant pools based on the admission prediction model that can help admission officers with structuring the admission process.

\subsection{Context}
The case university is one of many selective institutions in the United States where standardized test scores such as the SAT have been used in the admission process to prioritize the review order of applications. Like many other colleges, the case institution began test-optional undergraduate admissions during the COVID-19 pandemic which provided an emergent need for an alternative to the SAT-based heuristic.


\begin{figure}
\begin{center}
\centerline{\includegraphics[width=\columnwidth]{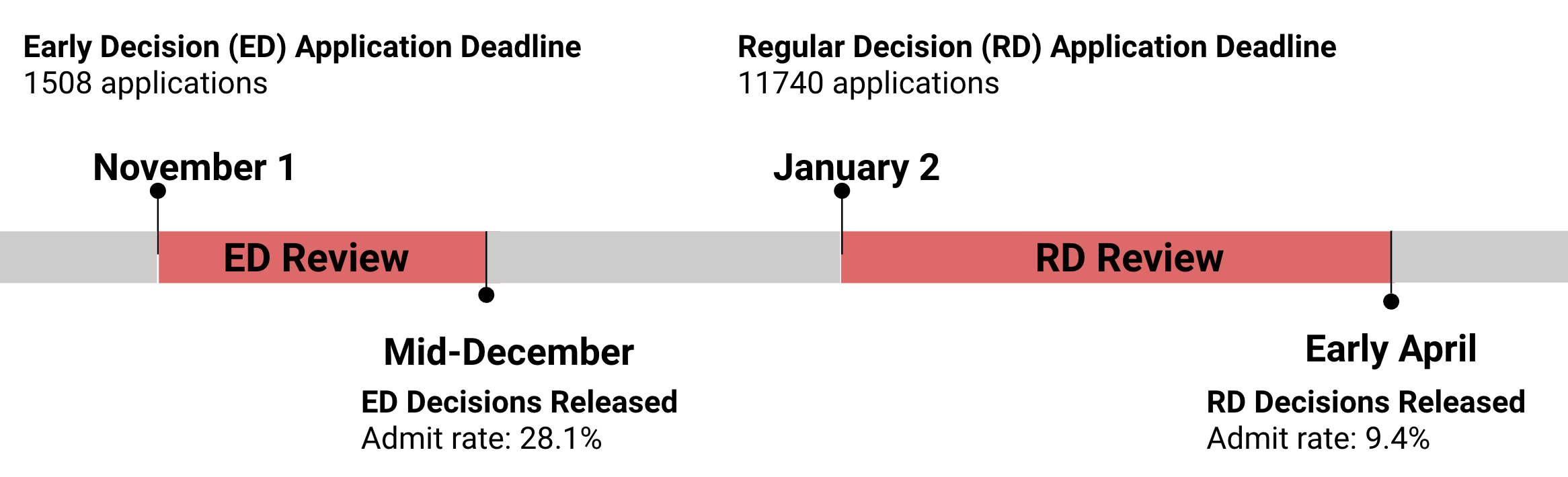}}
\caption{The 2019-2020 admission timeline for early and regular decisions at the case institution.} 
\label{timeline}
\end{center}
\end{figure}

Admission at the case institution begins with applicants submitting their applications for either the Early Decision (ED) or the Regular Decision (RD) admission cycle (see Figure~\ref{timeline}). In the 2019-20 cycle, 1,508 ED applications were received by November 1, and admission decisions were released in mid-December. During the RD admission cycle, 11,740 applications were submitted by January 2, and decisions were sent out by early April. The admit rates for ED and RD were 28.1\% and 9.4\%, respectively.\footnote{As an ED applicant, applicants apply to only one institution and must enroll at that school if admitted. Because of this binding nature, ED applicant pools usually have a higher rate of admission relative to RD applicant pools.} Application review was performed by professional admission staff in the case institution’s undergraduate admission office. Each admission officer was responsible for reviewing the applications from certain geographical regions, and a larger team of external reviewers supported their initial review of the applications. The review process spanned about one month for ED and about three months for RD.

The institution requires all first-year applicants to apply through the Common Application portal. Each completed application via Common Application includes a variety of information such as a personal essay (responding to one of seven prompts), descriptions of extracurricular activities, honors and awards, AP and/or IB scores, SAT or ACT scores, SAT subject scores, TOEFL/IELTS scores, and two college-specific essays. In addition, recommendation letters from two teachers, secondary school reports that include a recommendation letter from the guidance counselor, school profile and the official transcript of the applicant, and the current mid-year grade report are submitted alongside the Common Application. 

Like many other selective institutions in the US, the case institution has relied on standardized test scores, such as the SAT and the ACT, to organize the applicant pool for review. In particular, the admissions office at the case institution organized all applications into three pools, the ``Top’’, ``Middle’’, and ``Bottom’’ pool, using the SAT-based model. Starting with the Top pool, student applications were reviewed by admission staff and seasonal readers. Every application was reviewed by human reviewers regardless of their initial pool assignment, and admission staff reviewed applications in an iterative process to fill the annual admit target (i.e., the number of spaces available for matriculation). Grouping similar applicants together is a strategy that supports an equitable review process, since many highly qualified applicants apply to selective colleges like the case institution, but the human resources and the size of the first-year class are limited.

Indeed, we observe that the SAT-based segmentation into three pools shows a correlation with the admission outcome: 83.1\% of the final admitted class for the 2019-2020 admission cycle were in the Top pool, which comprised 56.6\% of the entire applicant pool. However, it is evident that the SAT-based approach did not replace holistic human review, as only 28.5\% and 8.8\% of applicants in the Top pool were identified as female and underrepresented minority (URM) applicants, respectively, whereas 51.4\% were female and 30.0\% were URM applicants in the final admitted class. In this work, we explore an admission prediction model that could replace the SAT-based heuristic that had been used at the case institution to triage the applicant pool. This new prediction model aims to align the formation of applicant pools with the holistic human review that follows. 

\subsection{Dataset description}
The dataset we used to build the prediction model comprises the student application data submitted to the case institution during the 2019-20 admission cycle as well as their final admission outcomes. First-year applicants apply through the Common Application which contains a fixed set of data fields including SAT and ACT scores, SAT subjects, Advanced Placement (AP) International Baccalaureate (IB), English proficiency test scores (TOEFL/IELTS), high school GPA, class rank, high school type (e.g., boarding school), intended major, legacy status, career interests, languages spoken, personal essay, application information (whether the application is for the ED or RD admission cycle, application fee waivers, etc.), extracurricular activities and time commitment, courses taken in the current year, high school disciplinary records, honors and awards, and several demographic indicators such as gender, ethnic background, citizenship, age, first-generation status, and parental levels of education. We consider all information presented in the Common Application except for personally identifiable information, such as first and last names, addresses, contact information, and names of high schools. 

In the 2019-20 admission cycle, all applicants were still required to report their SAT or ACT scores, and international applicants from non-English speaking countries were also required to submit their TOEFL or IELTS scores. In building the admission prediction model, however, we choose to remove SAT/ACT and TOEFL/IELTS scores from our feature set in order to simulate the test-optional admission policy. We note that the admission prediction model still includes the SAT subject scores, as they were optional for both test-required and test-optional admissions. We did not have access to certain application materials that are part of the Common Application but processed separately; as a result, several crucial pieces to application review such as college-specific essays, teacher and guidance counselor recommendation letters, high school reports, and transcripts are omitted in the feature set.  

We filter out a small number of duplicate applications. We also remove student-athletes and Reserve Officers' Training Corps (ROTC) applications who were recruited to the university prior to application submission. We impute missing values with a unique placeholder value and add an indicator variable for missingness. Categorical features are one-hot encoded, while some categorical values that occur with less than 1\% frequency are merged together as ``RARE". Descriptions of applicants' extracurricular activities and honors, which applicants type into a text box, are converted into TF-IDF features using both unigrams and bigrams. The final processed data has 13,248 rows and 1,434 columns, where each row corresponds to a single completed application submitted to the case institution for either the ED or RD admission cycle. Among the resulting features (columns), 888 are one-hot encoded categorical features, 171 are numerical features, and the remaining 375 are text features.

\subsection{Modeling from past admissions}
We model college admission prediction as a probabilistic binary classification problem and focus on the following two admission outcomes: admitted (including admitted and conditionally admitted applicants) and denied (including denied, wait-listed, and withdrawn applicants). We do not consider admitted applicants' matriculation decisions. We leave out a randomly sampled 20\% of the dataset for testing ($n=2650$), and train the model using the remaining 80\% of the dataset. We note that 11.5\% of applicants in the training data and 11.7\% in the testing data were granted final admission from the case institution in the 2019-20 admission cycle. We fit a Gradient Boosting Decision Trees model using \textit{scikit-learn}'s GradientBoostingClassifier using its default parameter settings \cite{scikit-learn}.

\begin{figure}
\begin{center}
\centerline{\includegraphics[width=\columnwidth]{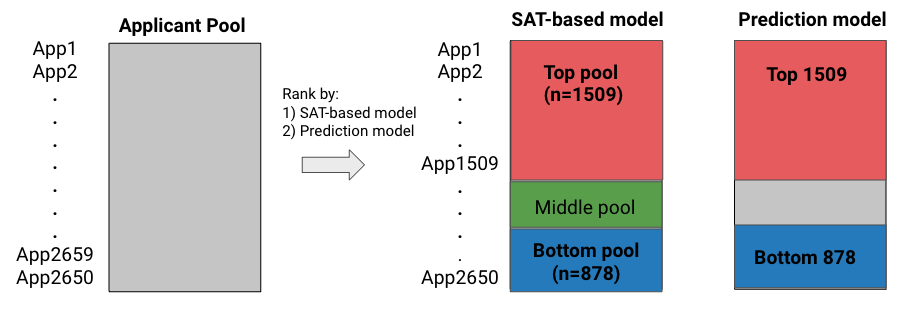}}
\caption{Illustration of the SAT-based and the prediction-based ``Top pool'' in the testing data ($n=2650$). The initial applicant pool is shown on the left. The two figures on the right represent the applicant pools organized by the SAT-based model (middle) and the prediction model (right). In the SAT-based model, 1,509 applicants are grouped into the Top pool. In the prediction model, the top 1,509 applicants as sorted by their predicted probability of admission are grouped into the prediction model's notion of the ``Top pool''.} 
\label{evaluation}
\end{center}
\end{figure}

We assess the potential of the prediction model to replace and improve the institution's SAT-based triaging process by using the following evaluation strategy. In the testing data, 1,509 applicants (57.0\% of the testing data) were placed in the Top pool by the SAT-based model, and 82.5\% of the final admitted class was identified from the Top pool. Similarly, we identify the top 1,509 applications based on the predicted probability of admission from the admission prediction model in order to simulate the model's version of the Top pool. We then compare the proportion of the eventually admitted class identified in the Top pools between the SAT-based model and the prediction model. We also compare the proportion of URM, female, and legacy applicants represented in the SAT-based and prediction-based Top pools to the proportion in the final admitted class (see Figure~\ref{evaluation} for a visual illustration). 

We note that the admission prediction model includes sensitive socio-demographic attributes of applicants (e.g., gender, race, and ethnicity) and excludes standardized test scores required for admission (i.e., SAT/ACT and TOEFL/IELTS scores) while including optional test scores (i.e. SAT subject scores). In Section \ref{sec:ablation}, we perform an ablation study of the prediction model to investigate the effects of standardized test scores and applicants' sensitive attributes on the model behavior. Specifically, we explore how much predictive value can be added to the model by including the required standardized test scores, and how much is lost by excluding SAT subject scores and the following sensitive socio-demographic attributes: race/ethnicity, sex, gender identity, URM status, first-generation status, parents' education level, religious preference, family income, age, citizenship, and applicants' geographical regions. We compare the proportions of the admitted class identified by the following four models. All four models are variants of the same baseline model. The baseline model itself does not use the required standardized test scores but includes sensitive attributes. Beyond the baseline model, we also evaluate a model that excludes the SAT subject scores from the baseline model, a model that includes the required standardized test scores in the baseline model, and a model that excludes sensitive attributes from the baseline model. 

\subsection{Construction of new applicant pools}

\begin{figure}
\begin{center}
\centerline{\includegraphics[width=\columnwidth]{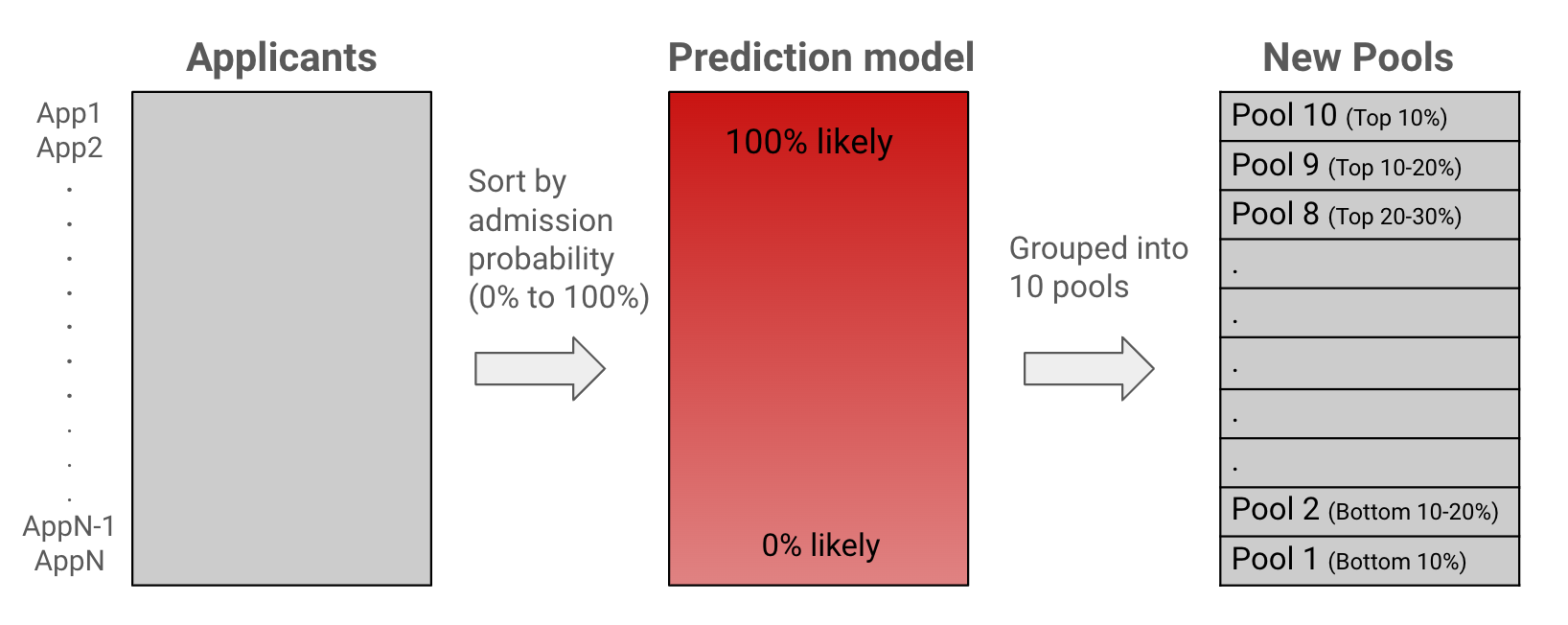}}
\caption{Construction of the new pools based on the admission prediction model. All applicants (left) are sorted by the predicted probability of admission (middle). The applicants with the top 10\% admission probability are placed in Pool 10, the next 10\% in Pool 9, and so on (right), to produce 10 different applicant pools with varying predicted acceptance rates.} 
\label{pools}
\end{center}
\end{figure}

While the SAT-based model used by the case institution defined only three pools (Top, Middle, and Bottom), the prediction model offers a straightforward way to create more fine-grained pools and convey their probabilistic semantics. One approach to creating fine-grained pools is to sort applicants by their predicted probabilities of admission and aggregate them into 10 different pools as follows: applicants with the top 10\% admission probability are placed in Pool 10, the next top 10\% is placed in Pool 9, and so on (see Figure~\ref{pools} for a visual illustration). Pool 10 is then predicted to have the highest number of admitted applicants, Pool 9 is predicted to have the second-highest number of admitted applicants, and so on. 

Note that this will not surface predicted probabilities of admission for individual applicants to the admissions officers, but only the assignment to pools. One can now convey to the admission officers that the density of qualified candidates in each pool is different, but that all pools contain qualified and not qualified candidates. To set expectations, one can also communicate the percentage of applicants predicted to be admitted (i.e., predicted admit density) in each pool, as it may not vary linearly across the pools (e.g., pool 10 may contain 50\% predicted admits, while pool 9 contains only 20\%). 

This use of machine-predicted probabilities to guide decision-making in admission is not meant to make judgments about individuals; rather, it is creating pools of applicants with different admission rates. In particular, we conjecture that much of the uncertainty captured by the prediction model is epistemic in nature, not aleatoric \cite{hullermeier_aleatoric_2021}. This means that the uncertainty comes from the model's limitation to truly understand an applicant's qualification, not from some external randomness. In the extreme, for a pool with a 90\% acceptance rate, 9 out of 10 applicants in that pool may individually have a 100\% probability of getting admitted, while one student has a 0\% probability. The probability therefore only makes sense if we think about pools of applicants instead of individual applicants. We thus stress that it is important to present results to human admission staff in terms of pools, not in terms of predicted probabilities of admission for individuals, to avoid any misconception of what the model is capable of.

In Section \ref{sec:uncertainty}, we assess the calibration of the predicted admission rates for each pool by checking whether or not the predicted admission rate matches the actual admission rate in each pool. In addition to analyzing the prediction uncertainty in the aggregate pools of applicants, we further explore the uncertainty of the prediction model by examining the distribution of the individual predicted probabilities for denied and admitted applicants.
\section{Results}

We evaluate the potential of the admission prediction model as a replacement for and an improvement over the SAT-based approach previously used at the case institution. We find that the prediction model outperforms the SAT-based model by placing more admits in the Top pool and fewer admits in the Bottom pool. The Top pool of the prediction model also more closely matches the final admitted class in terms of the female, URM, and legacy composition than the SAT-based Top pool. We analyze both the aggregate and individual predicted probabilities of the admission prediction model and find that the new applicant pools constructed from the prediction model are well-calibrated by comparing the predicted and the actual admit rates in each pool.

\subsection{Comparison to the SAT-based model}
\begin{figure}
\begin{center}
\centerline{\includegraphics[width=\columnwidth]{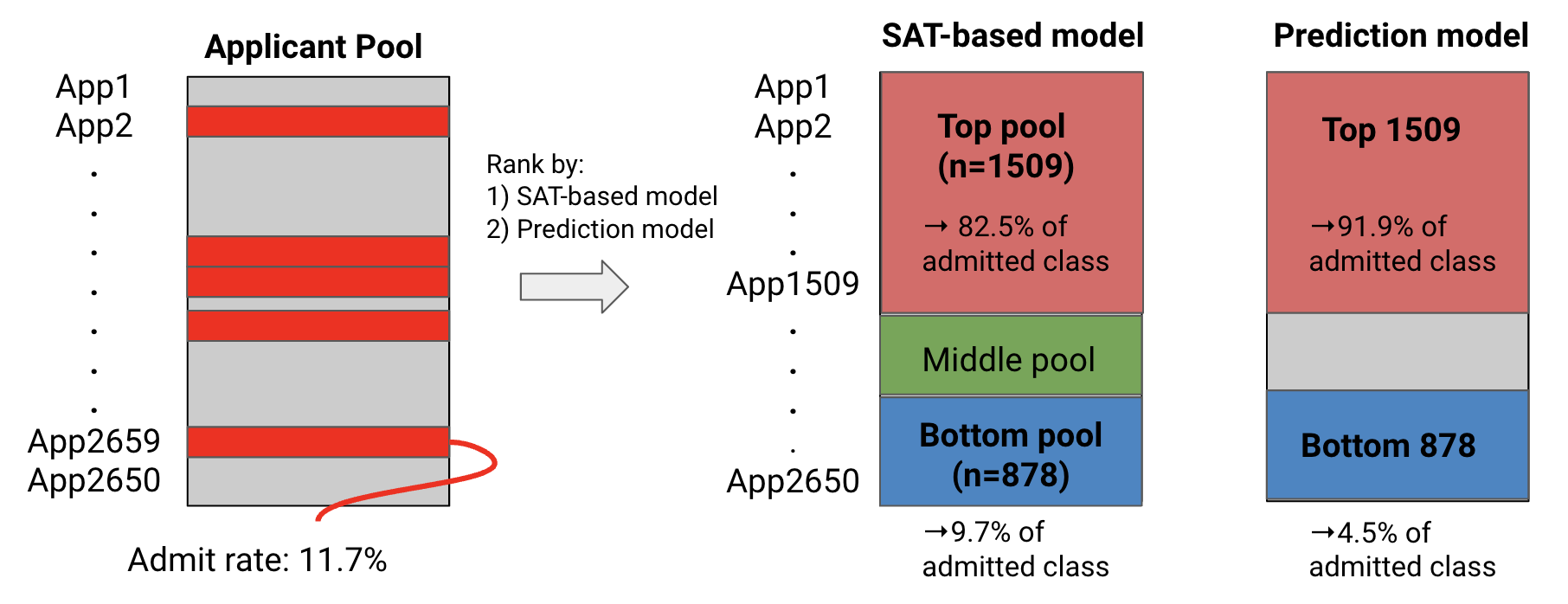}}
\caption{Illustration of the proportion of admitted class identified by the SAT-based model (middle) and by the prediction model (right).} 
\label{results}
\end{center}
\end{figure}

As shown in Figure~\ref{results}, we find that the prediction model outperforms the SAT-based model in having more admits in the Top pool and fewer admits in the Bottom pool. Specifically, the Top pool of the prediction model identifies 91.9\% of the final admitted class, compared to the 82.5\% in the SAT-based Top pool ($N=309$, $\chi^2(1) = 12.206$, $p<0.001$). Conversely, the Bottom pool of the prediction model consists of 4.5\% of the admitted class, while the SAT-based Bottom pool consists of 9.7\% of the admitted class ($N=309$, $\chi^2(1) = 6.2642$, $p=0.006$). These results show that the prediction model is significantly better aligned with the admission criteria of the case institution than the SAT-based model.

\begin{figure}
\begin{center}
\centerline{\includegraphics[width=\columnwidth]{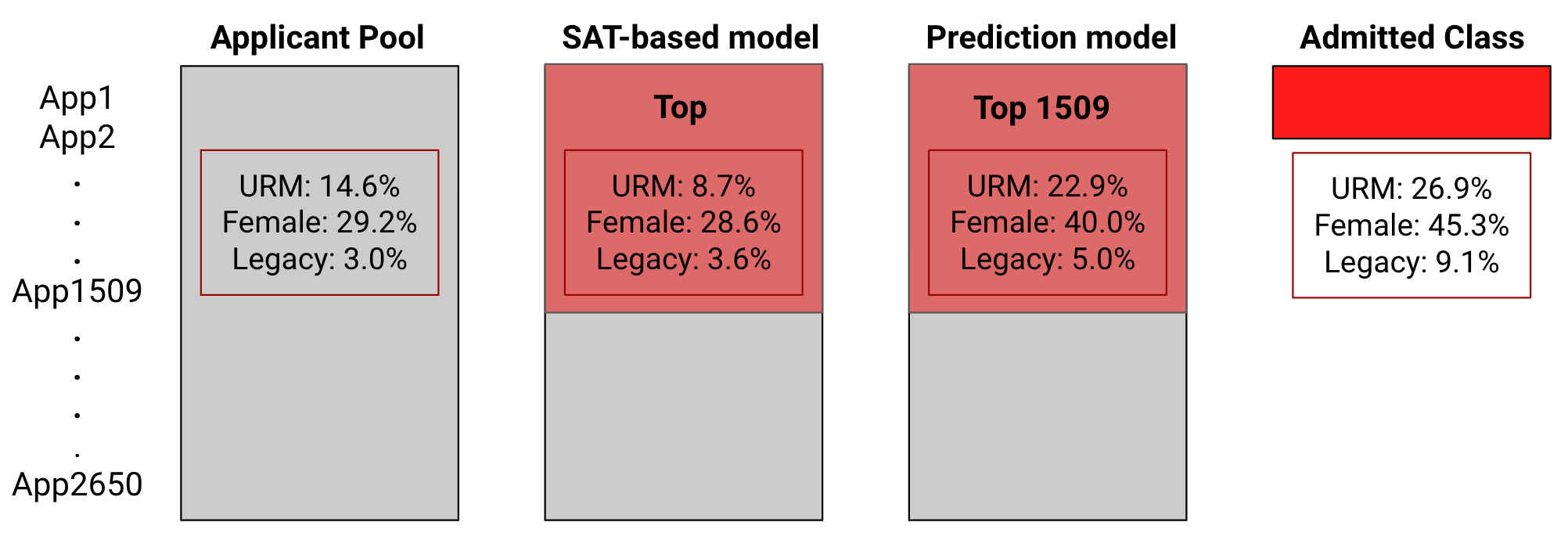}}
\caption{Proportion of URM, female, and legacy applicants in each of the following pools (from left to right): 1) the entire applicant pool ($n=2650$), 2) the Top pool of the SAT-based model ($n=1509$), 3) the Top pool of the prediction model ($n=1509$), and 4) the final admitted class ($n=309$).} 
\label{makeup}
\end{center}
\end{figure}

In addition, Figure~\ref{makeup} shows that the Top pool of the prediction model more closely matches the final admitted class in terms of the female, URM, and legacy distributions than the SAT-based Top pool. The URM, female, and legacy student proportion in the overall applicant pool is 14.6\%, 29.2\%, and 3.0\%, respectively. We see that 8.7\%, 28.6\%, and 3.6\% of the SAT-based Top pool are URM, female, and legacy applicants, while 22.9\%, 40.0\%, and 5.0\% of the prediction-based Top pool are URM, female, and legacy applicants. The prediction model thus produces a Top pool that more closely matches the demographic makeup of the final admitted class, where 26.9\% of admits are URM applicants, 45.3\% are female applicants, and 9.1\% are legacy applicants.


\subsection{Ablation Study}\label{sec:ablation}
\begin{figure}
\begin{center}
\centerline{\includegraphics[width=\columnwidth]{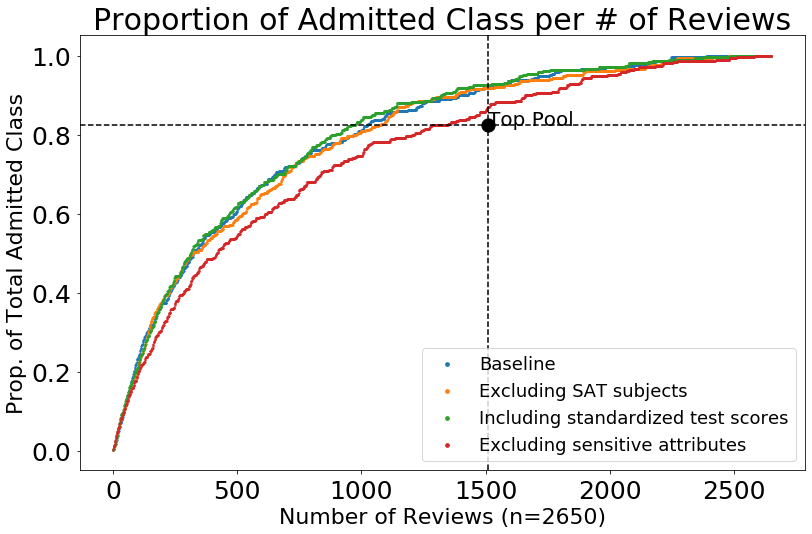}}
\caption{The proportion of total admitted class captured for each number of applications reviewed in the testing data ($n= 2650$) for the baseline model (blue), the model that excludes SAT subject scores (orange), the model that includes standardized test scores (green), and the model that excludes sensitive attributes (red).} 
\label{ablation}
\end{center}
\end{figure}


We find that the model that excludes the SAT subject scores achieves the same performance as the baseline model, both identifying 91.9\% of the admitted class in their respective Top Pools. We also find that the model that includes standardized test scores has comparable performance to the baseline model that excludes them ($N=309$, $\chi^2(1) = 0.090$, $p=0.3819$). However, the model that excludes sensitive attributes performs worse than the baseline model: it identifies 87.1\% of the total admitted class in the testing set compared to the 91.9\% in the baseline prediction model ($N=309$, $\chi^2(1) = 3.8684$, $p=0.0246$). In fact, we find that the model without sensitive attributes (and without standardized test scores) shows similar performance to the SAT-based model ($N=309$, $\chi^2(1) = 2.4592$, $p=0.0584$).

Figure~\ref{ablation} compares the predictive performance for these four models in terms of the proportion of applicants in the admitted class given a number of reviewed applicants, where applicants are reviewed in the order of their predicted scores (from highest to lowest). The dashed vertical line indicates the number of applicants in the SAT-based Top pool, with its corresponding predictive performance (dashed vertical line). We use this visualization to further understand the added predictive value of standardized test scores and sensitive socio-demographic attributes in the admission prediction model at all stages of a structured reviewing process. For any number of reviewed applicants, the baseline admission prediction model (excluding standardized test scores) is able to identify a similar proportion of admitted applicants as the model that includes standardized test scores as well as the model that excludes the SAT subject scores. The model that excludes sensitive attributes, however, performs consistently lower in identifying applicants in the admitted class at any stage of the review process. In fact, it performs similarly to the SAT-based model (indicated by a black point) for the number of reviews corresponding to the SAT-based Top pool size.


\begin{table}
\centering
\caption{Model comparison of the proportion of the admitted class, URM, female, and legacy applicants in each model's Top pool ($n=1509$): SAT-based model; baseline prediction model; baseline with SAT subject scores removed; baseline with standardized test scores added, and baseline with sensitive socio-demographic attributes removed. Full applicant pool ($n=2650$) characteristics are presented as a point of reference.}~\label{tab:ablation-makeup}
\small
\begin{tabular}{lcccc}
\toprule
& \textbf{Admitted} & \textbf{URM} & \textbf{Female} & \textbf{Legacy} \\
 \midrule
Applicant pool & --- & 14.6\% & 29.2\% & 3.0\% \\
\midrule
SAT-based model & 82.5\% & 8.7\% & 28.6\% & 3.6\% \\
Baseline model  & 91.9\% & 22.9\% & 40.0\% & 5.0\% \\
\quad Remove SAT subject scores & 91.9\% & 23.5\% & 40.0\% & 5.2\% \\
\quad Add standardized test scores & 92.6\% & 20.3\% & 40.8\% & 5.0\% \\
\quad Remove sensitive attributes & 87.1\% & 14.6\% & 31.5\% & 5.0\% \\
\bottomrule
\end{tabular}
\end{table}

We further examine how the four models compare in terms of the proportion of URM, female, and legacy students that they predict to be in the Top pool. Table~\ref{tab:ablation-makeup} shows that there are no significant differences in the composition of the Top pool as a result of adding standardized test scores to the baseline model, though it reduces the proportion of URM applicants by 2.6 percentage points (pp) ($\chi^2=2.98, p=0.085$). However, removing sensitive socio-demographic attributes from the baseline model not only reduces the recall rate as noted above, but it also significantly reduces the proportion of URM applicants by 8.3pp ($\chi^2=34.0, p<0.001$) and female applicants by 8.5pp ($\chi^2=23.6, p<0.001$) in the Top pool. The share of legacy applicants in the Top pool remains unchanged from adding or removing these features.

\subsection{Evaluating uncertainty of the prediction model} \label{sec:uncertainty}


\begin{figure}
\begin{center}
\centerline{\includegraphics[width=\columnwidth]{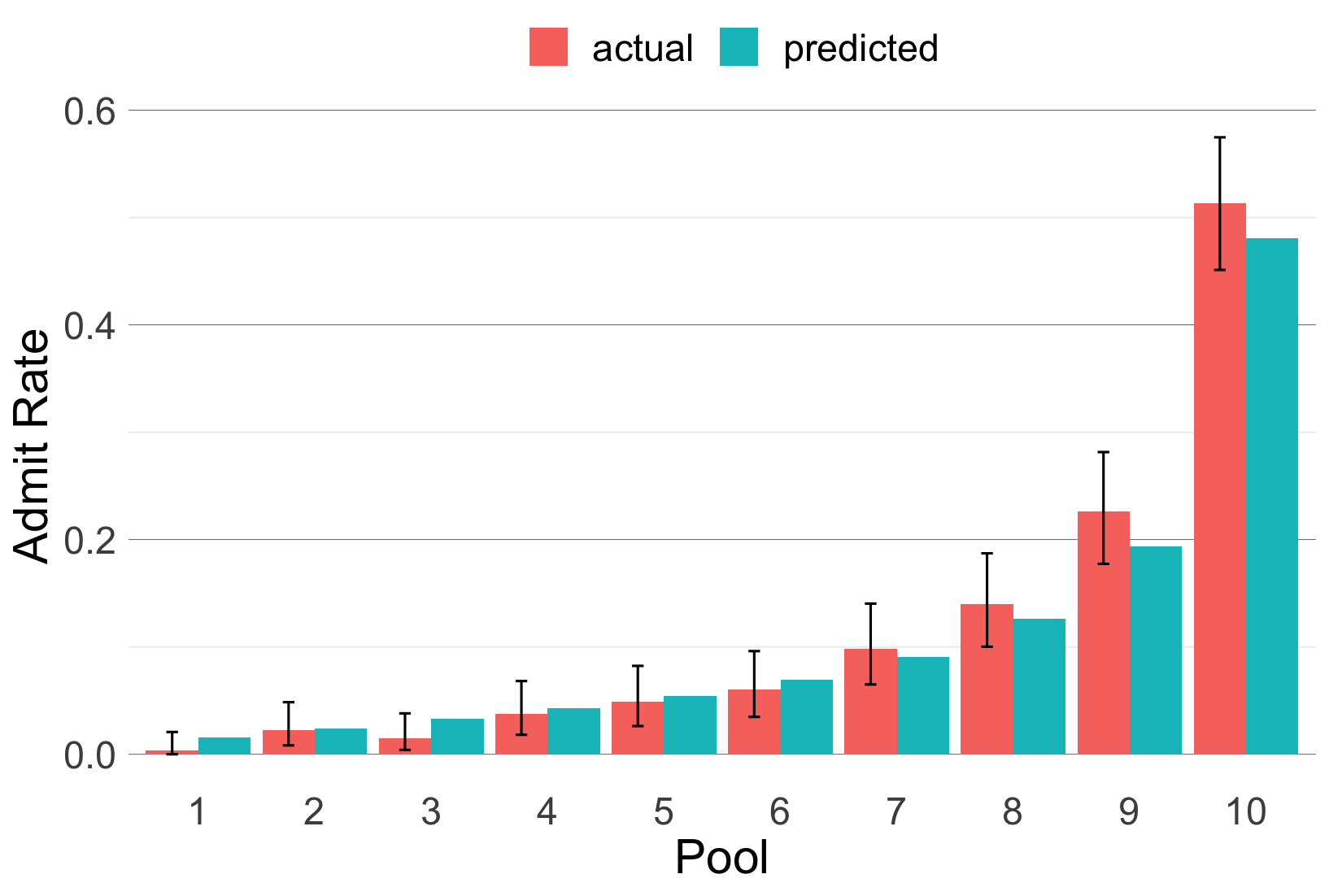}}
\caption{Predicted admit rate (blue) and actual admit rate (red) for each of the 10 pools constructed from the prediction model. Error bars indicate 95\% confidence intervals computed using the Clopper-Pearson (exact) method.} 
\label{calibration}
\end{center}
\end{figure}

        
\begin{figure*}
    \centering
    \includegraphics[width=0.43\linewidth, height=0.29\textwidth]{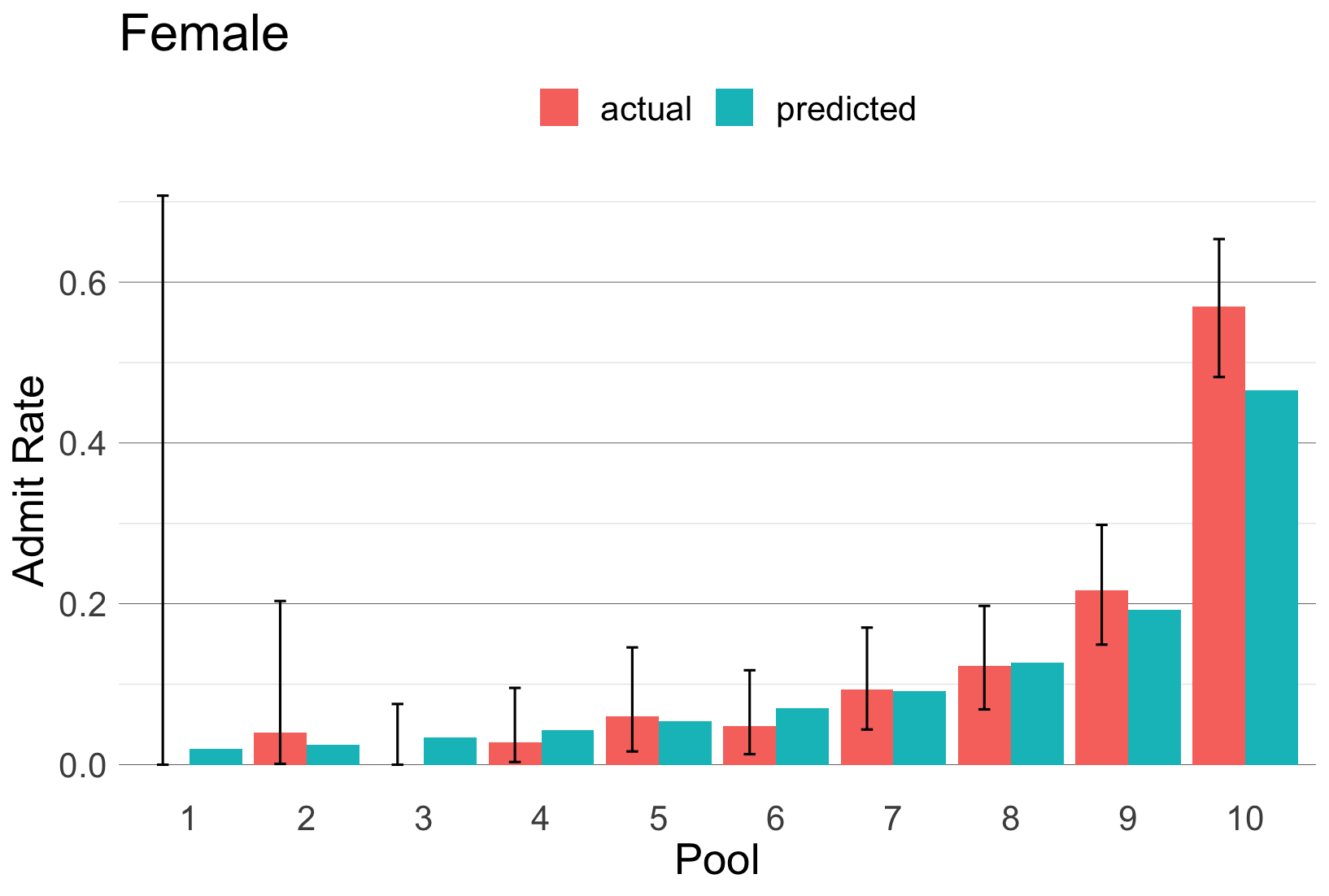}
    \includegraphics[width=0.43\linewidth, height=0.29\textwidth]{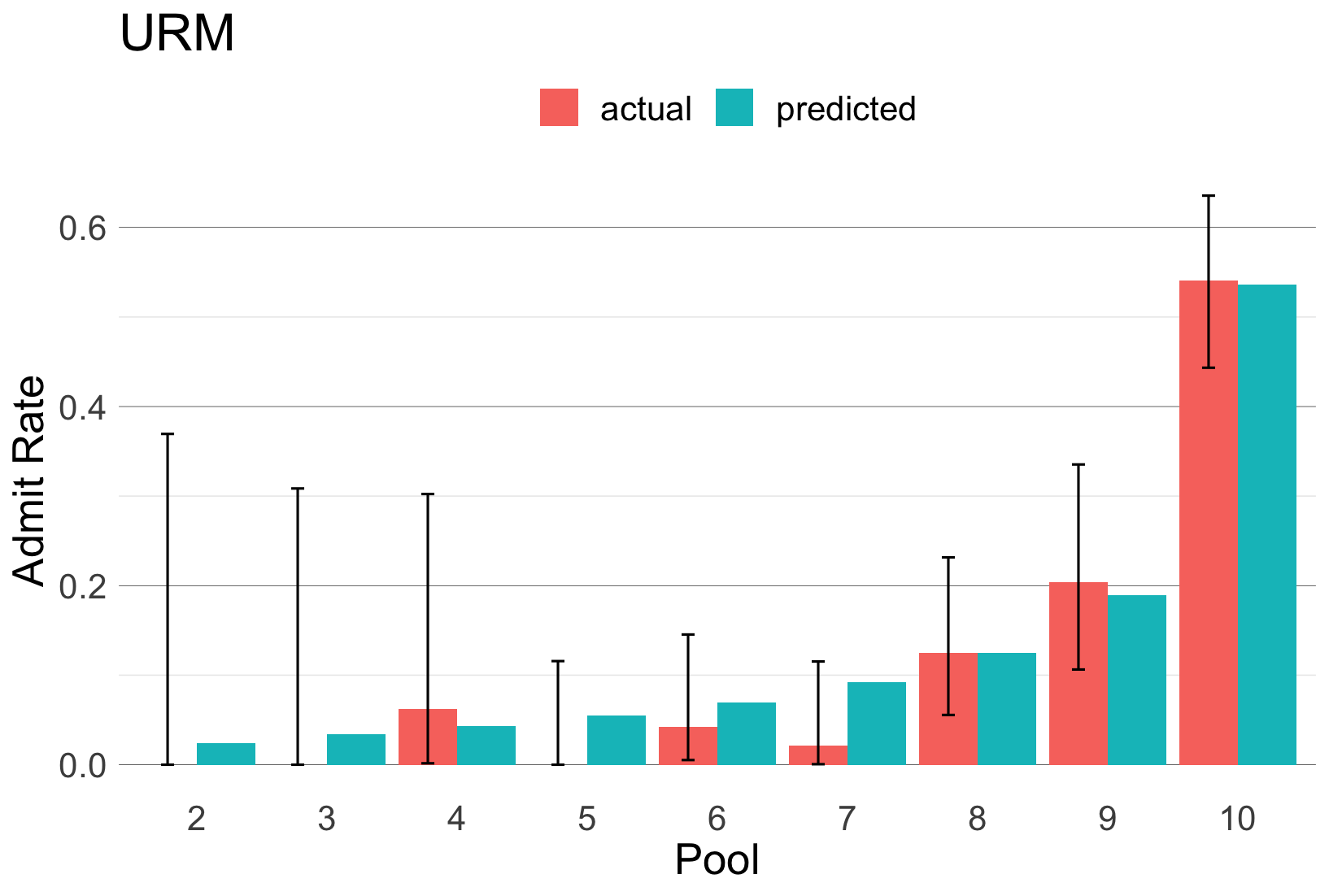}
    \caption{Predicted admit rate (blue) and actual admit rate (red) of female (left) and URM (right) applicants for each of the 10 pools constructed from the prediction model. Error bars indicate 95\% confidence intervals computed using the Clopper-Pearson (exact) method.}
    \label{calibration_group}
\end{figure*}

We find that new applicant pools constructed with the predicted probabilities are reasonably well-calibrated at the level of the defined pools. We compute the predicted admission rate by taking the average of the predicted probabilities in each pool, and the actual admission rate by taking the proportion of actual admits in each pool. As presented in Figure~\ref{calibration}, we see that the predicted admission rate closely follows the actual admission rate in each pool ($r(10)=0.998$, $p < 0.001$). Figure~\ref{calibration_group} also shows a similar trend for female applicants ($r(10)=0.996$, $p < 0.001$) as well as URM applicants ($r(10)=0.985$, $p < 0.001$) across pools.

\begin{figure}
\begin{center}
\centerline{\includegraphics[width=\columnwidth]{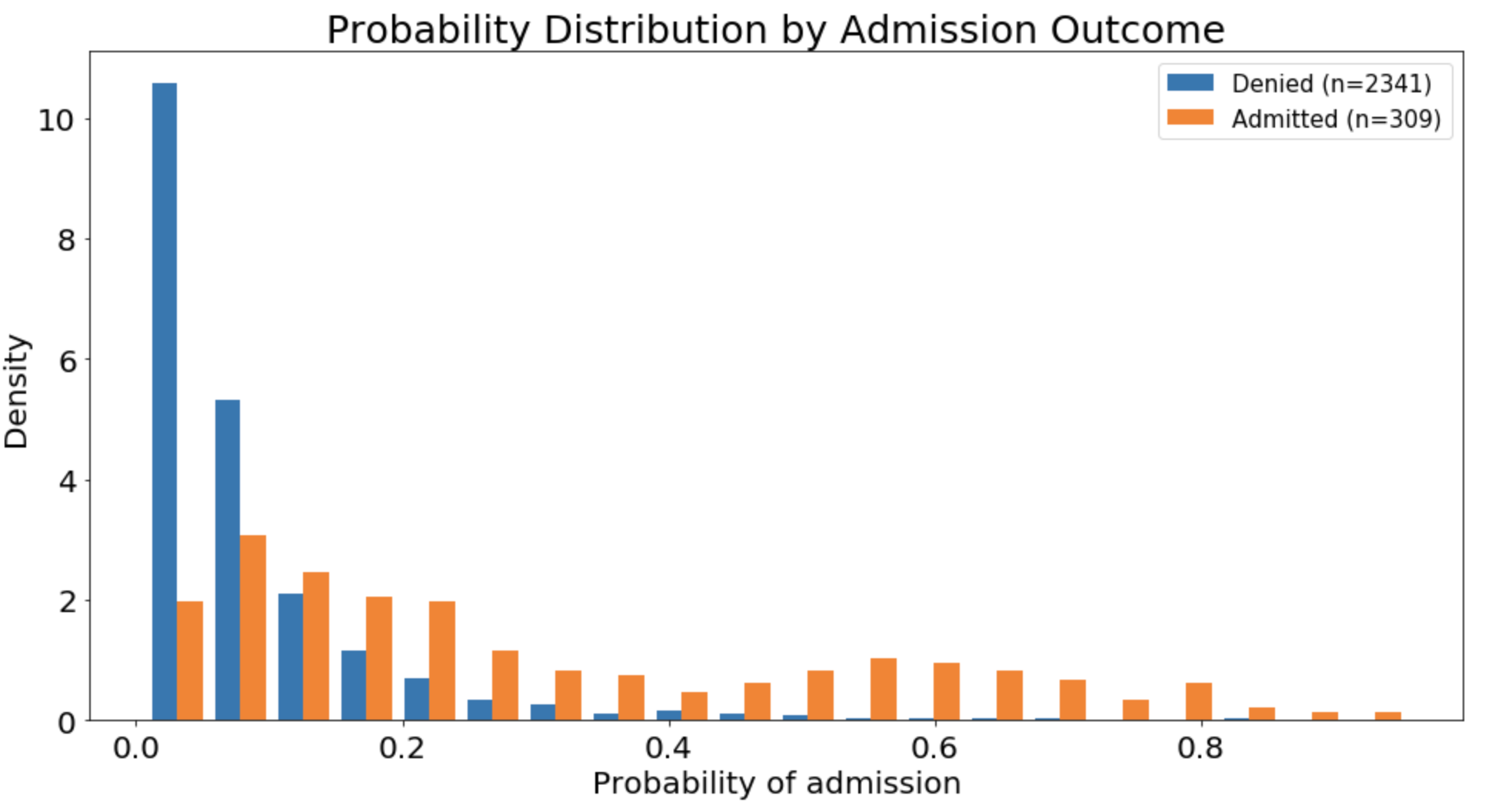}}
\caption{Density plot of estimated admission probability scores from 0.0 to 1.0 for denied (blue) and admitted (orange) applicants in the testing set.} 
\label{density}
\end{center}
\end{figure}

While the model is reasonably calibrated across the ten pools, an applicant's placement in a given pool does not indicate whether they should be admitted or not. For example, a student in Pool 10 is not necessarily a stronger applicant than a student in Pool 9. We highlight this issue by examining the distribution of predicted probabilities for individual applicants. Consider the following distribution of predicted probabilities for denied and admitted applicants shown in Figure~\ref{density}. We see that most of the predicted probabilities are concentrated on the lower end for the denied applicants, suggesting that the prediction model is able to accurately discern most of the denied applicants. On the other hand, the predicted probabilities for the admitted applicants are more widely spread across the entire range of predicted probabilities, implying that the prediction model has limited knowledge about who gets admitted. This suggests that the final admission decisions may depend largely on human evaluation of other parts of the applications such as teacher recommendations, personal essays, and transcripts that are missing from the prediction model; in other words, the prediction model exhibits significant epistemic uncertainty \cite{hullermeier_aleatoric_2021}. The aggregate pools are meant to provide a coarse yet meaningful organizational structure of the application pool for effective reading for admission officers, and as a safety net to minimize the chance that any qualified applicant gets overlooked.


\section{Discussion}
In this work, we set out to answer the following question: How well can a prediction model trained on past admission data replace and improve on the traditional SAT-based heuristic to organize the applicant pool for review? We identify three aspects that contribute to answering this question. First, we found that the prediction model outperforms the SAT-based model by placing more admits in the Top pool and fewer admits in the Bottom pool. Second, the Top pool of the prediction model more closely matches the final admitted class in terms of the female, URM, and legacy composition than the SAT-based Top pool. Finally, we showed that the new applicant pools constructed from the prediction model are reasonably calibrated by comparing the predicted and the actual acceptance rates in each pool, allowing for direct interpretation of the prediction model's applicant pools in a holistic admission process.

\subsection{Implications}
Many institutions began test-optional admission in response to testing site closures during the COVID-19 pandemic, which made the traditional SAT-based heuristic impractical. In our development and evaluation of the admission prediction model, we found that the prediction model may serve as a practical alternative to the SAT-based model for organizing the applicant pool; it can be trained using other already available student information in the Common Application excluding the standardized test scores (SAT and ACT) and English proficiency scores (TOEFL and IELTS). Because the prediction model represents a larger set of information provided in the student application, it can also be seen as more holistic compared to the SAT-based model which is only based on standardized test scores and a few demographic markers. 

Comparative analysis of the prediction-based and the SAT-based Top pools suggests that the prediction model may not only replace but also improve the triaging process in important ways. We found that in the past application cycle, the prediction model would have identified 9.4\% more admitted applicants in the Top pool and 5.2\% fewer admitted applicants in the Bottom pool compared to the SAT-based model. In other words, the prediction model outperforms the SAT-based model in terms of both the true positive rate (i.e. more admits in the Top pool) and the false negative rate (i.e. fewer admits in the Bottom pool). The prediction model further improves on the traditional SAT-based model as the prediction-based Top pool contains a pool of applicants with a female, URM, and legacy composition that is closer to the actual admitted class compared to the SAT-based Top pool. These results suggest that the prediction model is able to organize the applicant pool in a way that better reflects the institution's admission goals, which is in line with the fact the prediction model is indeed trained on the institution's past admission decisions. 

The results of the ablation study provide valuable insights into the relative importance of different features in developing the admission prediction model. We found that the inclusion of the required standardized test scores or exclusion of SAT subject test scores do not significantly affect the performance of the prediction model, yet excluding sensitive socio-demographic attributes has a negative impact in terms of both the proportion of the admitted class and the demographic composition of the Top pool. These results support the use of sensitive attributes in developing admission prediction models to improve both accuracy (i.e. proportion of admitted class identified in the Top pool) and equity (i.e. demographic composition of the Top pool compared to that of the admitted class). Our findings contribute to the ongoing discussions about the ethical and practical implications of the use of sensitive attributes in predictive modeling in education \cite{kleinberg, yu20, yu21, baker}.

The upside of using the prediction model instead of the SAT-based model is that the admission practices are no longer tied to the biases in the standardized test scores that do not align with the institution's values \cite{freedle, doi:10.1177/003172170208400411, doi:10.1177/016146811311500406, rosser1989sat, gender_sat, doi:10.3102/0013189X18762105}. However, because the admission prediction model is directly informed by past admission decisions, it is important to ensure that the past admission data is appropriate for developing and using the resulting prediction model. So, any use of the model should be accompanied by a governance process that ensures that the dynamics of the admission process continue to reflect the potentially changing values of the institution. A prediction model is a tool, where thoughtful use could untether admission practices from biases in external scores to iteratively improve admission practices, but thoughtless use can also cement existing inequities \cite{grade_critique}.

A new affordance of the prediction model is that the new applicant pools come with predicted acceptance rates that are well-calibrated to the true acceptance rates. In contrast, the SAT-based model did not yield pools that are calibrated to actual acceptance rates. This suggests that the prediction model may further inform the admission process by enabling a direct interpretation of the applicant pools in terms of their predicted acceptance rates. For example, the institution may set itself the goal to double the acceptance rates in the lower pools, encouraging admission staff to find creative new ways of identifying qualified applicants that do not fit the typical profile. In this way, the prediction model may serve as a way to target a subset of applicants to examine more closely and highlight new applicants who might have been overlooked.


\subsection{Limitations and future work}
While the admissions prediction model presented in this paper provides promising results, there are several limitations that need to be considered. First, admission decisions are not independent as they are about creating a class of admitted students, but they were modeled as if they were independent when building the prediction model. In practice, admitting a cellist, for example, can reduce the probability of admission for other cellists in the pool because the orchestra needs students who play each of the instruments. Future work could explore ways to incorporate this interdependence into the model to provide more accurate results. 

Second, we note that our subgroup analysis only considered a limited number of applicant subgroups, specifically female, URM, and legacy applicants. While our analysis provides valuable insights about the model behavior for some key subgroups, we recognize that it does not provide a comprehensive understanding of the prediction model's impact on specific racial and ethnic groups, and other important subgroups such as international, low-SES, and first-generation applicants. Future work should aim to identify other relevant applicant subgroups and carefully examine the prediction model on these groups as well as the intersectionality of these groups \cite{baker_algorithmic_2022,kizilcec2020algorithmic}. 

Third, the application data used to train the admission prediction model only included information that was readily available in the Common Application, which omits several important components of applications, such as college essays and recommendation letters. Future work should explore ways to incorporate natural language features into the prediction model to enable a more comprehensive evaluation of student applications. Recent studies have sought to quantify salient aspects of open-response text data such as extracurricular activity descriptions~\cite{park_inequality_2023} and the style and content of personal essays~\cite{doi:10.1126/sciadv.abi9031}, which provide a useful starting point.

Another potential direction for improving the prediction model is to account for covariate shifts. For example, the admission prediction model was developed on admission data where SAT/ACT scores were mandatory, which may have a different distribution from future years where tests are optional or no longer factor into the decision at all. Even though the SAT and ACT scores are excluded entirely from the feature set for building the admission prediction model, there may potentially be a large covariate shift between the year in which the model is trained and tested, and the year in which the trained model is to be deployed. The model evaluation results presented in this work are based on the testing set from the same year that the model is trained on, and this may not yield an accurate estimate of the model performance in another year because of the potential shift in empirical distributions. Future work may attempt to account for the potential covariate shift across datasets from different years by training a classifier for identifying whether a given data point belongs to the training or the testing distribution, and using it to assign more weights to training instances that are closer to a testing distribution~\cite{covariate_shift}.

Given the interaction of the prediction model with the human decision-making process, the full impact of the prediction model on the admission process and final decisions requires evaluation in a randomized controlled trial. In addition to assessing the impact of including the prediction model in the admission practice, future work could explore ways to provide explanations to admission staff as to how the pools are computed (i.e., algorithmic transparency) and why the prediction model places a given applicant in a particular pool (i.e., explainable AI). Comparing the admission decisions between the prediction model and admission staff may help human decision-makers reflect on their decisions to see where they may have blind spots, which may even challenge existing norms about what is valued in admitted students. It may also help improve the prediction model by identifying where the model may have blind spots, and admission staff may be able to offer insight into where additional data collection may be helpful to make the predictions better.

\section{Acknowledgments}

This research was supported in part by NSF Award IIS-2008139.
All content represents the opinion of the authors, which is not
necessarily shared or endorsed by their respective employers and/or
sponsors.


\bibliographystyle{ACM-Reference-Format}
\bibliography{sampleThesis}


\end{document}